\documentclass[aps,prb,superscriptaddress,amsmath,amssymb,twocolumn,amsfonts]{revtex4-1}
\usepackage{amsmath}
\usepackage{graphicx}
\usepackage{hyperref}
\usepackage{cleveref}
\usepackage{color}
\usepackage{float}
\usepackage{soul}
\usepackage{ulem}
\usepackage{braket}
\usepackage{appendix}

\newcommand\bea{\begin{eqnarray}}
\newcommand\eea{\end{eqnarray}}
\newcommand\beq{\begin{equation}}  
\newcommand\eeq{\end{equation}}

 \usepackage{physics}
\usepackage{tensor,upgreek,float}
\usepackage{parskip}
\begin{document}

\title{Spin-imbalance induced buried topological edge currents in Mott \& topological insulator heterostructures.}

\author{Rahul Ghosh}
\email{rahul.ghosl@niser.ac.in}
\affiliation{School of Physical Sciences, National Institute of Science Education and Research, Jatni 752050, India}
\affiliation{Homi Bhabha National Institute, Training School Complex, Anushaktinagar, Mumbai 400094, India}

\author{Subhajyoti Pal}
\affiliation{School of Physical Sciences, National Institute of Science Education and Research, Jatni 752050, India}
\affiliation{Homi Bhabha National Institute, Training School Complex, Anushaktinagar, Mumbai 400094, India}

\author{Kush Saha}
\email{kush.saha@niser.ac.in}
\affiliation{School of Physical Sciences, National Institute of Science Education and Research, Jatni 752050, India}
\affiliation{Homi Bhabha National Institute, Training School Complex, Anushaktinagar, Mumbai 400094, India}

\author{Anamitra Mukherjee}
\email{anamitra@niser.ac.in}
\affiliation{School of Physical Sciences, National Institute of Science Education and Research, Jatni 752050, India}
\affiliation{Homi Bhabha National Institute, Training School Complex, Anushaktinagar, Mumbai 400094, India}
\date{\today}

\begin{abstract}
We theoretically investigate the heterostructure between a ferrimagnetic Mott insulator and a time-reversal invariant topological band insulator on the two-dimensional Lieb lattice with periodic boundary conditions. Our Hartree-Fock and slave-rotor mean-field results incorporate long-range Coulomb interactions. We present charge and magnetic reconstructions at the two edges of the heterostructure and reveal how \textit{buried} topological edge modes adapt to these heterostructure edge reconstructions. In particular, we demonstrate that the interface magnetic field induces a spin imbalance in the edge modes while preserving their topological character and metallic nature. We show that this imbalance leads to topologically protected buried spin and charge currents. 
The inherent spin-momentum locking ensures that left and right movers contribute to the current at the two buried interfaces in opposite directions. We show that the magnitude of the spin-imbalance induced charge and spin current can be tuned by adjusting the spin-orbit coupling of the bulk topological insulator relative to the correlation strength of the bulk Mott insulator. Thus, our results demonstrate a controlled conversion of a spin Hall effect into an analog of a charge Hall effect driven by band topology and interaction effects. These topologically protected charge and spin currents pave the way for advances in low-energy electronics and spintronic devices.
\end{abstract}

\maketitle

\section{Introduction}

The need for sources of dissipationless current sources in electronic devices is ubiquitous. Stable spin current sources are essential for advancing spin-based electronics such as spin logic circuits to spin-based sensors and allow for increased energy efficiency \cite{application-spin}. Spin current sources are essential in manipulating spin information within qubits, the critical ingredient of quantum computers \cite{application-quant}. Similarly, charge currents that can flow without scattering are vital for low-energy electronics. However, the realization of dissipation-free current sources continues to face severe roadblocks. 
Perturbations such as spin relaxation, disorder and damping effects, temperature sensitivity, interface scattering, quantum tunneling and interference, and many-body effects are detrimental to reliable dissipationless spin and charge current sources at room temperature \cite{spin-curr-scatt}. Protection against such generic agencies is difficult to guarantee, given the wide range of energy and length scales involved in these perturbations.

In this regard, topologically protected edge modes in time-reversal symmetric (TRS) band topological insulators (TI) that support helical edge modes with spin momentum locking are obvious candidates for dissipationless spin currents. However, magnetic impurities at the edge of a \textit{time-reversal-invariant} (TRS) band-topological insulator (TI)\cite{PhysRevLett.95.146802, PhysRevLett.95.226801} can induce scattering between left and right movers due to the lifting of Kramer's degeneracy\cite{RevModPhys.82.3045}. The edges are expected to lose their topological protection in such a situation. Moreover, the helical edges only support dissipationless spin-currrents in the ideal case. It is unclear if TRS-TI can be tweaked to support topologically protected charge currents. Can the magnetic field-induced TRS-breaking be turned from a foe to a friend \cite{mag-topo-1, mag-topo-2}? 

The interplay of strong correlation-induced magnetism and topology from spin-orbit has thus been actively investigated. The interplay of interaction effect on BTI edge modes has been studied on the Bernevig-Hughes-Zhang (BHZ) model\cite{doi:10.1126/science.1133734}. Within Dynamical Mean Field Theory (DMFT)\cite{capone-0, capone-1} and considering interaction and spin-orbit coupling on the \textit{same region} of the lattice in a strip geometry, the existence of topologically protected metallic edge mode has been established in the paramagnetic regime. The BHZ TI model kept in proximity to a \textit{paramagnetic} Mott insulator (MI), has been shown to support metallic edge models along with interesting edge reconstructions\cite{buried-int}. Finally, paramagnetic DMFT solution on interacting square lattice with spin-orbit interactions sandwiched between two Mott insulators have described the nature of metallic state induced in the edge layers of the Mott insulators\cite{PhysRevB.87.161108}. 

However, the fate of the edge modes when the magnetic field is \textit{induced by proximity} to a magnetic Mott insulator retaining the heterostructure geometry and contending with the complications from spin and charge reconstructions has received little attention. Such a situation can be engineered by considering a heterostructure of a strong-correlation-driven magnetic Mott insulator and a band topological insulator (BTI). This situation allows studying magnetic field effects on `buried' helical edge modes.

\begin{figure}[t]
	\centering
	\includegraphics[width=0.99\linewidth]{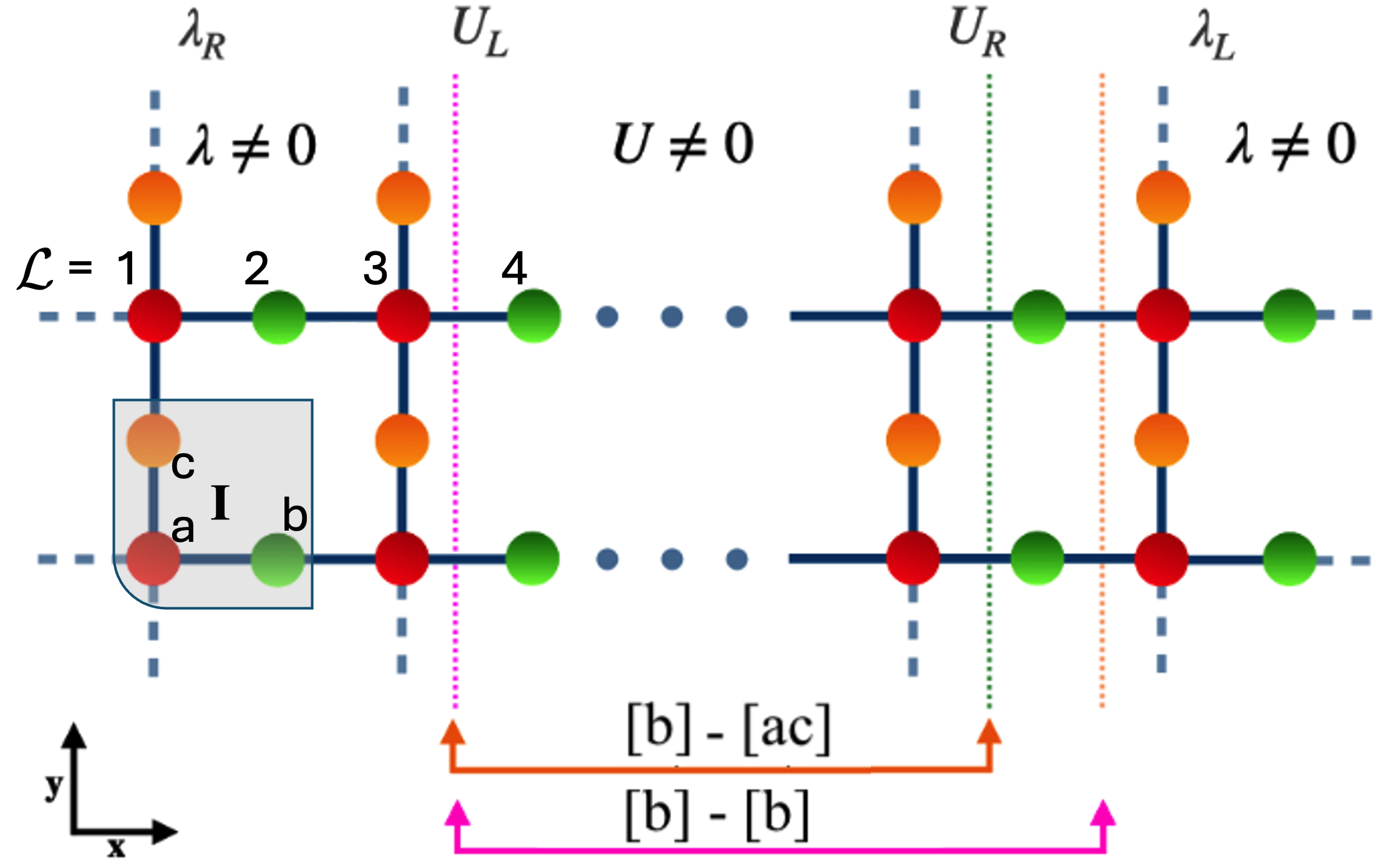}
	\caption{\textbf{Heterostructure schematic :} We divide a periodically identified Lieb lattice into  $U\neq 0$  and   $\lambda\neq0$ parts along the x-direction. The resulting interfaces (vertical dashed lines) are parallel to the y-direction.  Unit cells and sites of $\lambda_R-U_L$  interface ($I_1$) and $U_R-\lambda_L$  interface ($I_2$) are labeled as follows. 
 The [ac] edge denotes that the (ac) sites on the $U_L$ edge hybridize with the (b) sites on the region labeled by $\lambda_R$. At the [b] edge, the (b) sites on the $U$ side interface with (ac) sites of the TI. The three-unit cell consists of sites labeled by 'a, 'b', and 'c.' The two heterostructure cases with interfaces [b]-[ac] and  [b]-[b] are demarcated at the bottom with  arrows.  The three site unit cell is marked  with `I' and the vertical layers of (ac) sites and (b) sites demarcated by $\mathcal{L}=$1, 2 ,3 ... are also shown.}
	\label{fig:1}
\end{figure}

From a general standpoint, heterostructures of dissimilar materials have yielded several surprises. The most striking example has been the demonstration of high mobility two-dimensional electron gas\cite{mobility} superconductivity\cite{supersc} ferromagnetism\cite{ferro_supersc} and magnetoresistence\cite{magnetores}  in the heterostructure of LaAlO$_3$ and SrTiO$_3$, a band and a Mott insulator respectively. This has spawned a huge body of investigation in correlated heterostructures \cite{PhysRevB.70.241104,PhysRevB.85.235112,PhysRevB.81.115134,PhysRevB.81.115134,PhysRevLett.101.066802,PhysRevLett.108.246401,PhysRevB.84.085103}. In addition, junctions of superconductor-topological insulator\cite{RevModPhys.83.1057, PhysRevB.81.241310} and junctions between topological insulators\cite{PhysRevB.88.245120} are being actively being investigated.  

In this spirit, we consider the heterostructure of a TRS-TI and a ferrimagnetic Mott insulator to examine the fate of buried topological edge modes. However, such an investigation has to tackle spin, charge, and magnetic reconstructions as is well-known in conventional MI and band insulator (BI) \cite{rmp-hetero, PhysRevB.81.115134}.  
These reconstructions require modeling long-range Coulomb interaction along with local Hubbard-model-type interaction effects\cite{charge-reconstruction}. Treating such interaction effects by interfacing with spin-orbit coupled TI can help answer several outstanding questions. These include how edge modes form when the TI edges hybridize with the magnetic MI (unlike TI with open boundary conditions). How do the edge modes accommodate the induced magnetization arising from the above-mentioned heterostructure edge reconstruction? What is the nature of spin and charge transport? Finally, what is the impact of different kinds of edge termination on buried edge modes? In this paper, we address these questions within the following setup.

For our study, we consider an underlying Lieb lattice in two dimensions. The heterostructure is constructed by stabilizing the Mott insulator with Hubbard repulsion $U$ on all sites on one half of the lattice and a topological insulator with spin-orbit coupling  $\lambda$ on the other. In addition, we consider long-range Coulomb interaction over the entire lattice. We employ unrestricted Hartree Fock mean field theory (HF) that allows direct access to magnetism to show that the ferrimagnetic order in the Mott insulator induces a spin imbalance among the edge modes. We compute spin-resolved edge currents to show that \textit{the proximity-induced spin imbalance induces a topologically protected spin-polarized current}. We then demonstrate the tunability of the spin-polarized current with interaction and spin-orbit coupling. We support the mean-field conclusion of the existence of the edge modes for large interaction strengths using a strong interaction slave rotor (SR) mean-field approach\cite{PhysRevB.70.035114, PhysRevB.76.195101} where we also incorporate long-range Coulomb interaction effects.

We report several significant findings. We demonstrate the charge and magnetic edge reconstruction at the heterostructure interface with different kinds of junction geometries.
We establish that the magnetic field at the interface induces a spin imbalance in the edge modes while preserving the helical nature of the metallic edge modes. We show that the spin imbalance leads to the conversion of the pure spin current into a topologically protected charge current. We also demonstrate that the magnitude of these induced spin and charge currents can be tuned by controlling the spin-orbit coupling of the bulk topological insulator relative to the correlation strength of the bulk Mott insulator. Through this work, we thus demonstrate a controlled conversion of the spin Hall effect into an analog of the charge Hall effect, driven by band topology and interaction effects. 

The paper is organized as follows: In Sec. II we present the heterostructure model, method and observables we compute. We discuss the results in Sec. III. We conclude the paper in Sec. IV.

\section{Model, method \& Observables}  \label{sec_2}
The Hamiltonian for the heterostructure defined on the Lieb lattice, Fig.~\ref{fig:1}, is as follows:

\begin{equation}
\begin{aligned}
&\mathcal{H}=-t \sum_{\langle i, j\rangle \in \Lambda} \sum_{\sigma} (d_{i \sigma}^{\dagger} d_{j \sigma} +h.c.)\\&+i \lambda \sum_{\langle\langle i, j\rangle\rangle\in \Lambda_{\lambda}} \sum_{\sigma, \sigma'}\nu_{i j} (d_{i\sigma}^{\dagger} \sigma_{\sigma, \sigma'}^{z} d_{j \sigma'} 
)+U \sum_{i\in \Lambda_{U}} n_{i \uparrow} n_{i \downarrow}.
\end{aligned}
\label{ham1}
\end{equation}

Here, $\Lambda\equiv\Lambda_U\oplus\Lambda_\lambda$ denotes the union of lattice points of whole heterostructure lattice with $\Lambda_U$ and $\Lambda_\lambda$ are the set of the lattice sites of the $U\neq0$ and $\lambda\neq0$ sides. $d_{i \sigma}^{\dagger}$ and $d_{i \sigma}$ are electronic creation and annihilation operators respectively for spin state $\sigma$. $t$  refers to nearest neighbor hopping amplitude, while $\lambda$ is the spin-orbit coupling involving next-nearest neighbors hopping. $\nu_{i j}=
\left(\mathbf{d}_{i j}^{1} \times \mathbf{d}_{i j}^{2}\right)_{z}
=\pm 1$ depending on the clockwise/anti-clockwise traversal of electrons around the 'a' sites. This fact is encoded using $d^1_{ij}$ and $d^2_{ij}$ are the two unit vectors along the two nearest-neighbor bonds connecting `a' sites to the `b' and `c' sites  respectively\cite{weeks}. Lastly, $\sigma^z$ denotes the third Pauli matrix in spin space with components $\sigma, \sigma' \in {\uparrow, \downarrow}$. We note that the spin-orbit coupling does not mix the spin species and should be considered a chirality-inducing term. The long-range Coulomb interaction between all the charges is taken into account via a self-consistent solution of the Coulomb potential. It is crucial to incorporate the long-range Coulomb interaction to control the amount of charge transferred across the interface. At a site $i$, the Coulomb potential is defined as $\phi^{i}$. The long-range Coulomb interaction Hamiltonian at a mean-field level as in literature \cite{PhysRevB.77.014441,PhysRevB.88.115136} is as follows:  

\begin{equation}
   \label{mad1}
    H_{LRC}=\sum_{i} \phi^{i}\hat{n}_{i}
\end{equation}
where $\hat{n}_i=\sum_\sigma d^\dagger_{i,\sigma}d_{j,\sigma}$ and $\phi^{i}=\sum_{j \neq i} a_M t \frac{\left\langle\hat{n}_{j}\right\rangle-z_{j}}{\left|\vec{r}_{i}-\vec{r}_{j}\right|}$ is the Coulomb potential at site $i$. In the definition of $\phi^i$, $\left\langle\hat{n}_{j}\right\rangle$ are electron charge densities, $j$ runs over all lattice sites and $a_M(=e^2/\epsilon a t)$, with $\epsilon$  and a denoting the dielectric constant and the lattice parameter, respectively. 
. The background charge is assumed to be $z_j=1$ for all sites. We assume a half-filled lattice with $\langle n_i\rangle=1$. We also employ two additional notations in the paper to analyze the results. Firstly, a unit-cell based label $\{I _\alpha\}$ where $I$ is the unit cell index and $\alpha$ refers to the atom $\alpha(=a,b,c)$ of the $I^{th}$ unit cell and secondly a layer-label $\mathcal{L}$. These are shown in Fig.~\ref{fig:1}. 

\textit{Hartree-Fock treatment:} The interaction term in Eq.~ \ref{ham1} is first treated within the HF mean field theory. The mean field self-consistency is carried out \textit{along the x-direction while translation invariance is assumed in the y-direction} (as defined in Fig.~\ref{fig:1}). This allows us to define the eigenstates as a function of $k_y$. To set up notation, we define the Hartree-Fock mean-field eigenvalues and eigenvectors as $\{\epsilon_{\kappa\sigma}(k_y)\}$ and $\{\langle I_\alpha,k_y,\sigma|\psi_{\kappa}\rangle\}$ respectively.  
We note that while translation invariance is imposed along the y direction, $\{\langle I_\alpha,k_y,\sigma|\psi_{\kappa}\rangle\}$ can vary with the unit cell index along x. We also construct 'layer-projected eigenvectors' labeled by the index $\mathcal{L}$ as indicated in Fig.~\ref{fig:1}.  


The mean field solution of Hamiltonian Eq. \ref{ham1} is discussed in Appendix A.1. The mean field solution is used in the calculation of the Coulomb potential as presented in Appendix A.2. Finally, in Appendix A.3,  we provide details of the mean-field self-consistency and iterations.

From the HF calculations we compute the unit-cell resolved density of states (DOS) $N_I(\omega)=\sum_{\kappa,k_y,\alpha\in I}|{\langle I_\alpha,k_y,\sigma|\psi_{\kappa}\rangle}|^2\delta(\omega-\epsilon_{\kappa,\sigma}(k_y))$, where $I$ refers to unit-cell index along the x-direction, as discussed above. We also compute the spin resolved unit cell dependent band-dispersions, $\epsilon_{{I},\sigma}(k_y,\omega)=\sum_{\kappa}|\sum_{\alpha\in I}{\langle I_\alpha,k_y,\sigma|\psi_{\kappa}\rangle}|^2\delta(\omega-\epsilon_{\kappa,\sigma}(k_y))$. 
To track unit cell dependence of magnetization, we compute the magnetization of the $I^{th}$ unit-cell  $M_z\equiv(1/6)\sum_{\alpha}(\langle n_{\alpha_I\uparrow}\rangle-\langle n_{\alpha_I\downarrow}\rangle)$ averaged over $\alpha=(a,b,c)$ sites in the $I^{th}$ unit cell. 

In Appendix A.4, we derive the expression of the charge current at a unit cell $I$ of the heterostructure. A brief discussion is given below. 
\begin{equation}
\begin{aligned}
    &\dot{n}_I=i[H,n_I]=-\nabla.J_I=-\sum_{\sigma=\uparrow\downarrow}(j_{\sigma}^{(I\rightarrow I+\delta_x)}+j_{\sigma}^{(I\rightarrow I-\delta_x)}\\&+j_{\sigma}^{(I\rightarrow I+\delta_y)}+j_{\sigma}^{(I\rightarrow I-\delta_y)})
\end{aligned}
\label{cur1}
\end{equation}
where $n_{I}=\sum_{\sigma,\alpha}n_{I\alpha\sigma}$ and $\hbar=1$. Now defining spin resolved current flowing in positive direction along `x' as $j^x_{\sigma}=\sum_{I}j_{\sigma}^{(I\rightarrow I+\delta_x)}$ and along positive `y' as $j^y_{\sigma}=\sum_{I}j_{\sigma}^{(I\rightarrow I+\delta_y)}$, we can express the total charge current flowing in $y$ direction is $ J \equiv j^y_{\uparrow}+j^y_{\downarrow}$. Similarly, $J_s=j^y_{\uparrow}-j^y_{\downarrow}$ defines the spin current along y.  \\
\textit{Slave-rotor treatment:} We use the standard SR strong-interaction mean-field approach to corroborate the Hartree-Fock results. While the SR approach cannot capture magnetic order, it allows calculation of the Mott gap and Mott to metal transition retaining charge fluctuations, which is missed in Hartree-Fock theory. In the present work, we carry out the SR approach in a non-uniform charge density background due to the Coulomb potential. The details of this calculation are given in Appendix B.1.
The main observable in our SR calculation is the unit-cell resolved density of states $N_{I}(\omega)$ defined and extracted from the SR single-particle Green's function, as derived in Appendix B.2. 

\section{Results} 
  We present results for  two kinds of heterostructure [b]-[ac]  and [b]-[b] as shown in Fig.~\ref{fig:1}. We first discuss the unrestricted Hartree-Fock (HF) mean field results and then discuss the (SR) results.

\begin{figure}[t]
	\centering
	\includegraphics[width=0.99\linewidth]{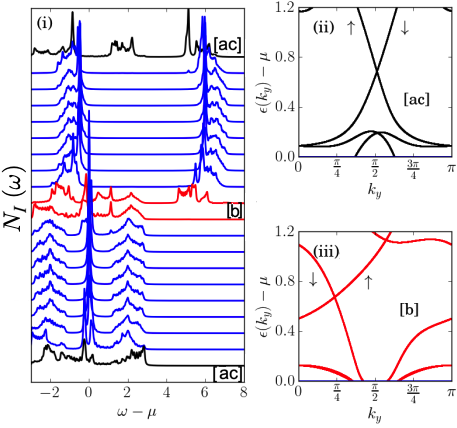}
	\caption{\textbf{Density of states \& edge modes for [b]-[ac] heterostructure}. (i) show the unit-cell projected DOS.  The interface unit cell  DOS (black) and (red), respectively, are for the  [b] and [ac] interfaces, one unit cell on either side of the interface. (ii) and (iii) show the spin-resolved momentum ($k_y$)-dependent buried-interface modes at the two interfaces. The dispersions are computed, including contributions from one unit cell on the $U\neq0$ and three on the $\lambda\neq0$ side, for the [ac] (black) and [b] (red) interfaces respectively. The results are presented for $\lambda=0.3t$ and $U=7t$ on a Lieb lattice with 20 three-site unit cells along the x-direction, while translation symmetry is assumed along the y-direction.}
	\label{fig:2}
\end{figure}
\subsection{Unit cell-resolved DOS \& buried topological edge modes}
Fig.~\ref{fig:2} (i) shows the unit-cell resolved density of states  (DOS) for [b]-[ac] heterostructure from the HF calculations. The unit-cell-resolved  DOS is plotted from the bottom (the terminating unit cell of the $\lambda\neq0$ region at the  $U_R-\lambda_L$ or [ac] interface) to the middle with two red unit cells one on either side of the  $\lambda_R-U_L$ or [b] interface, and going up to the top (the remaining terminating unit cell of the $\lambda\neq0$ region at the  $U_R-\lambda_L$ or [ac] interface).
The DOS of the interface unit cells colored in red and black are of particular interest to us. The DOS of the remaining unit cells are shown in blue. In the plot, $\omega-\mu=0$ demarcates the global chemical potential. For the bulk $U\neq0$ unit cells, the DOS shows a bulk gap $\sim U/t$ and the lower and upper Hubbard sub-bands as expected \cite{Lieb-1, Lieb-2}. For the $\lambda\neq0$ bulk, we recover the Lieb lattice spectrum supporting two dispersive bands and a flat band, with a separation of $4\lambda/t$ from both the dispersive bands\cite{weeks}. The DOS of the red and black unit cells both show finite spectral weight spread over in the energy regime where the respective bulk DOS have gaps. The spin-resolved band dispersion along $k_y$ arising out of the unit cells corresponding to the red  ([b] edge) and black ([ac] edge) DOS of Fig.~\ref{fig:2} (i)  are shown in Fig.~\ref{fig:2} (ii) and  Fig.~\ref{fig:2} (iii). Comparison with Fig.~\ref{fig:2} (i) shows that the edge modes are constructed from the spectral weight in the energy windows of the bulk gap seen for the red and black unit cells. 
We find clear spin-momentum locking, as is expected, at the edge of a topological band insulator. For clarity, we have only shown the edge mode above the Fermi energy; the same spin-momentum locking is seen for the edge modes below the Fermi energy.

At the edge of a conventional time-reversal symmetric band topological band insulator, the currents for the up and the down spin carriers are equal in magnitude, and the system supports helical spin currents. However, as a main result of the paper, we demonstrate that these buried edge modes are quite different from those in the conventional case. Due to the ferrimagnetic state of the Mott insulator, a spin imbalance is induced on the $\lambda\neq0$ edge unit cells, leading to a finite charge current instead of a pure spin current. To elucidate the physics, we first discuss the reconstruction of the charge and magnetism across the heterojunction.

\begin{figure}[t]
	\centering
	\includegraphics[width=0.99\linewidth]{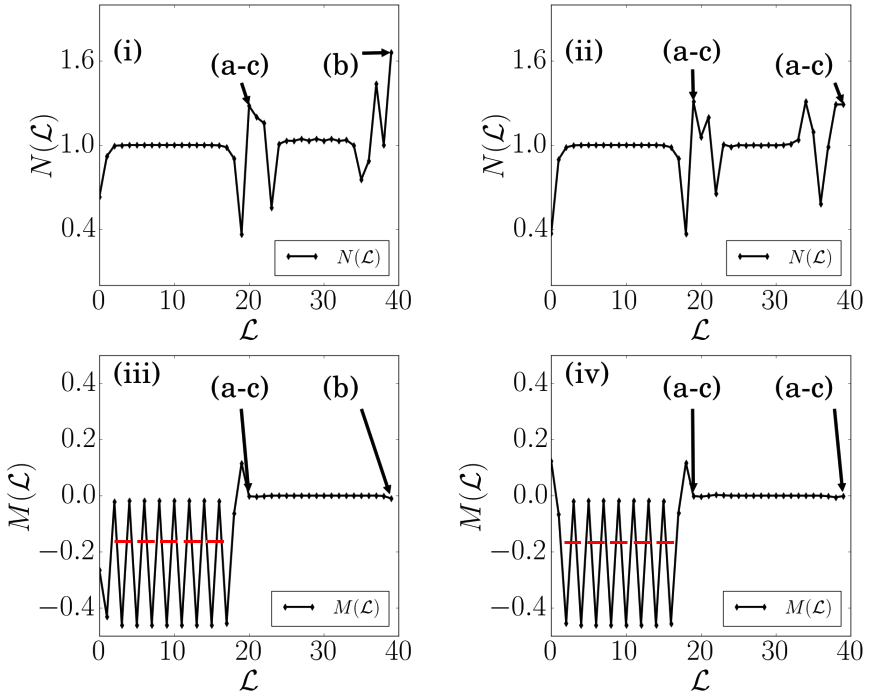}
	\caption{\textbf{Magnetic \& charge reconstructions:}  
 (i) and (iii) show the layer-dependent charge density $N(\mathcal{L})$ (black line) and  magnetization $M(\mathcal{L})$  profile along the x-direction perpendicular to the interface respectively for the  [b]-[ac] heterostructure. (ii) and (iv) show the corresponding results for the [b]-[b] heterostructure.  In all the panels alternate layers are composed of (a-c) and (b) sites along x- axis as can be seen from Fig.~\ref{fig:1}. The results are shown for $\lambda=0.3t$ and $U=7t$. }
	\label{fig:3}
\end{figure} 

\subsection{Edge reconstruction}
We first discuss the charge reconstruction as we traverse across the interfaces, shown in Fig.~\ref{fig:3} (i) for the [b]-[ac] heterostructure. Fig.\ref{fig:3} (i)  and (iii) show the layer ($\mathcal{L}$) resolved charge density profile (black curve) and magnetization for a fixed $U(=7t)$ and $\lambda(=0.3t)$ respectively. In Fig.~\ref{fig:3} (i) along the x-axis,  alternate layers are composed of (a-c) and (b) sites as can be seen from Fig.~\ref{fig:1}. The (a-c) and (b) sites of the $\lambda\neq0$ region interfacing with the $U\neq0$ region are marked by arrows in Fig.~\ref{fig:3} (i). The density and the magnetization are averaged over all (a-c) sites and (b) sites in each layer. 

From Fig.~\ref{fig:3} (i), we find that the average density per site in the bulk $U\neq0$ and bulk $\lambda\neq0$ is 1.  
The half-filling triggers a Mott insulating state in the bulk layer. The Mott state in the bulk $U\neq0$ region is characterized by the Mott-lobes separated by $U$ as seen in the  $U\neq0$ DOS in Fig.~\ref{fig:2} (i). Similarly, the $\lambda\neq0$ bulk behaves as a half-filled Lieb lattice.
We note substantial charge-density reconstruction at the interface that causes the filling to deviate locally from $\langle n_i\rangle=1$. As a result, the edge states are partially filled. For the (a-c) and (b) layers of the $\lambda\neq0$ side interfacing with the $U\neq0$ side, the average charge densities are larger than 1 that is compensated by strong suppression of change density of the interfacing layers of the $U\neq0$ side. \textit{The charge reconstruction induces a self-doping of the interface layers contributing  to the buried metallic edge modes between a Mott and a TRS topological insulator.}

From Fig.~\ref{fig:3} (iii), we find an alternating magnetization profile in the Mott regime with a non-zero average value (dashed horizontal line) as expected from the Lieb's theorem\cite{Lieb-0}, thus stabilizing a ferrimagnetic Mott insulator. We note that for the layers with (a-c) sites, the magnetization is depicted as an average value, while for the layers with (b)-sites, the magnetization is provided for the (b) site only. The net non-zero magnetization is computed by average unit cell magnetization ($M_z$), defined earlier. The same convention is followed for the $\lambda\neq0$ side. The bulk-magnetization of the $\lambda\neq0$ side is zero. However, for the unit cells on the left [b] and right [ac] interfaces, the average  unit cell magnetization for the $\lambda\neq0$ side are -0.0026 and -0.0043, respectively. In the following subsection, we discuss the importance of this small negative magnetization on the edge currents.

In  Fig.\ref{fig:3} (ii)  and (iv), we show the results for the [b]-[b] heterostructure. While the broad phenomenology remains the same, we see the change in the charge and magnetization reconstruction when the [b] edge is replaced by [ac] edge at layers $\mathcal{L}=0$ and 39.

\begin{figure}[t]
	\centering
	\includegraphics[width=0.99\linewidth]{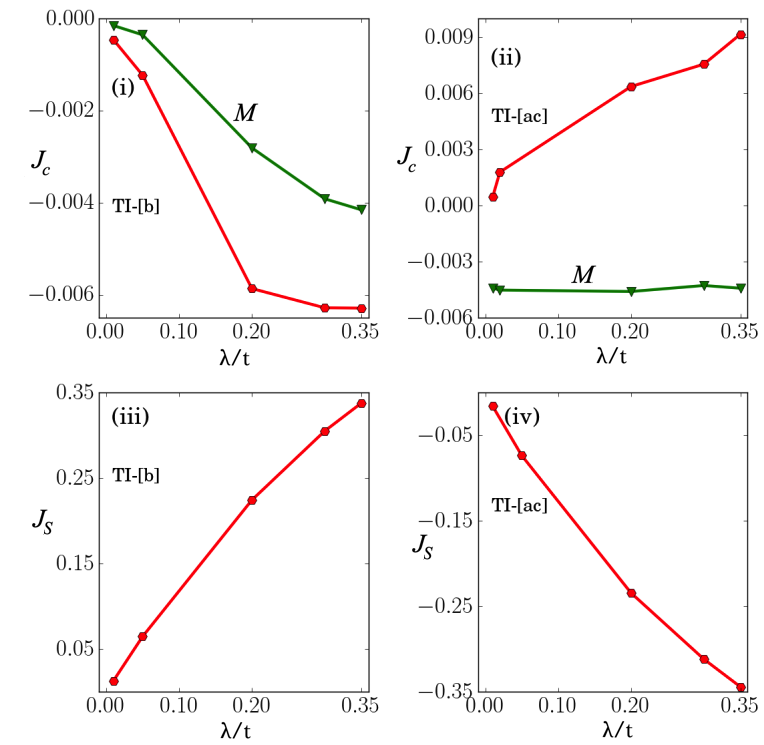}
	\caption{\textbf{Charge and spin currents for the [b]-[ac] hetereostructure: } 
(i) shows the charge current $J_c=J\uparrow+J_\downarrow$ (red circles) and magnetization  $M$ (green triangles) as a function of $\lambda$.  The magnetization $M$ is averaged over unit cell magnetization $M_z$ of three edge unit cells on the $\lambda\neq 0$ region. The left (right) panel shows the results for the [b] ([ac]) edges for $U=7.0t$. The corresponding spin currents ($J_s=J_\uparrow-J_\downarrow$) at the two edges are shown in panels (iii) and (iv). Both $J_c$ and $J_s$ are also averaged over contributions  of three edge unit cells on the TI side.}
	\label{fig:4}
\end{figure}

\subsection{Charge \& spin currents:}
We show the edge currents and magnetization for the [b]-[ac] heterostructure in Fig.~\ref{fig:4} for various $\lambda$ values for the two edges on the topological side. Numerically, for the calculation parameters, the charge current ($J_c$), the spin current ($J_s$), and the magnetization ($M_z$) penetrate up to three unit cells at the interface on the TI side (along the x direction). Hence, the results are shown by averaging $J_c$,  $J_s$ and $M_z$ over these three unit-cells. Similarly, averaged quantities are shown for the [ac] edge. 

Fig.~\ref{fig:4} panels (i) and (ii) respectively show the charge current ($J_c$) and  $M$ for the [b] and [ac] edges on the topological side. For all the $\lambda$ values, the induced magnetization at the TI edge is negative. The negative magnetization induces an increased down-spin occupation of the edge states at both [b] and [ac] edges. However, from Fig.~\ref{fig:2} (iii) and (ii), we find that the spin-momentum locking implies that the down spin electrons contribute to current along the negative y-direction for the [b] edge. In contrast, it contributes to current along positive y-direction at the [ac] edge. We emphasize that the edge modes in Fig.~\ref{fig:2} (iii) and (ii) are also computed, including contributions from three edge unit cells. For zero magnetization of the edge (for example, a TRS TI on the Lieb lattice interfacing with vacuum, the edge currents in opposite directions would be equal in magnitude. That would imply a net zero $J_c$, but a finite spin current equal to $J_s$. The charge current is clearly finite in the presence of the induced negative magnetization. It flows in the negative y-direction for the [b] edge and the positive y-direction for the [ac] edge. In addition, the edges support finite $J_s$.  Fig.~\ref{fig:4} (iii) and (iv) show the spin current ($J_s$) as a function of $\lambda$ for $U=7t$ for the [b] and the[ac] edge,s respectively. As expected, the spin current magnitudes for both edges increase with $\lambda$. However, unlike for a TRS-TI interfacing with vacuum, a part of the spin current is converted to the charge current $J_c$ due to the induced magnetic field at the edge. We emphasize that the full heterostructure is characterized by a common spin quantization axis, and only a weak magnetization is induced at the edge modes. Thus, the spin-momentum-locked helical nature of the edge states and the ensuing topological protection survive even in the presence of TRS breaking at the edge. Hence, we have shown a mechanism of converting a spin current into a charge current, which is equivalent \textit{to converting a spin-Hall effect to a charge Hall effect analog driven by band topology}.

\begin{figure}[t]
	\centering
	\includegraphics[width=0.9\linewidth]{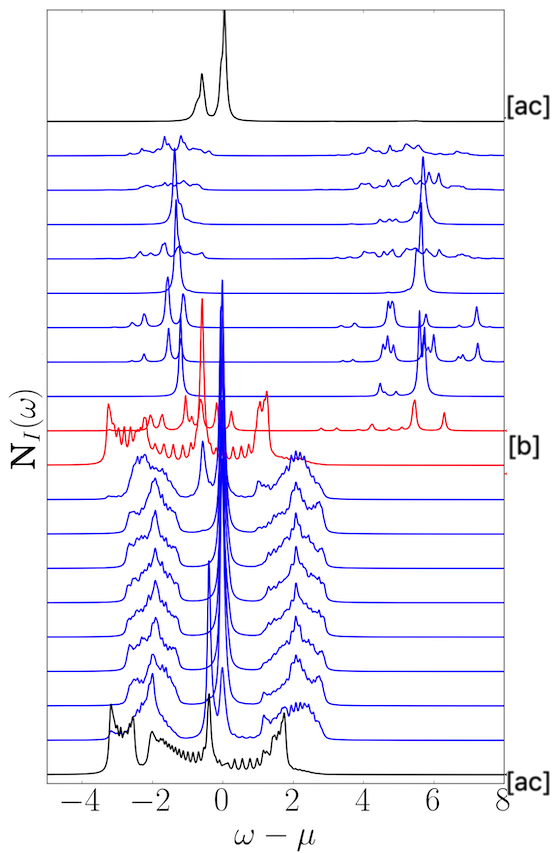}
	\caption{\textbf{Slave-rotor single-particle unit-cell resolved density of states for the [b]-[ac] heterostructure: } We show the unit-cell projected SR DOS. The interface unit-cell DOS (black) and (red), respectively, are for the  [b] and [ac] interfaces, one unit cell on either side of the interface. The results are presented for $\lambda=0.3t$ and $U=7t$ on a Lieb lattice with 20 three-site unit cells along the x-direction, while translation symmetry is assumed along the y-direction.}
	\label{fig:SR1}
\end{figure}
 
\subsection{Slave-rotor results}
The advantage is that SR  mean field theory retains charge fluctuations that are not captured in HF theory. However being a paramagnetic approach, it can only capture Mott-metal transition and misses out on magnetic order, as is true for the paramagnetic DMFT calculations. Thus, the SR is used here to demonstrate the validity of the HF result of the existence of the edge modes despite the large $U$ interfacing with the TI. The SR calculations, as discussed in Appendix B, are performed on the background of the Coulomb potential. 

Fig.~\ref{fig:SR1} shows the layer resolved DOS from SR mean-field theory for the [b]-[ac] heterostructure. The results presented are analogous to the HF results in Fig.~\ref{fig:2} (a). The bulk of the $U\neq0$ region supports a many-body Mott gap. Note that the sharp nature of the upper and lower Hubbard sub-bands arises from a cluster treatment of the rotor Hamiltonian, as discussed in Appendix B. The bulk $\lambda\neq0$ region shows TRS-TI band structure on the Lieb lattice. Finally, the interface layers for both the [b] and the [ac] edges show spectral weights that interpolate the bulk bands, indicating the same phenomena as in the HF theory. The appearance of the sharp peak at the Fermi energy in the interface unit-cells of the Mott regime is the well-known mid-gap resonance\cite{doped-mott-dmft} of doped Mott insulators. They arise out of Kondo screening of gapless spin states of Mott insulators by the carriers of the helical edges that penetrate the first unit cell of the MI at these two interfaces. Such mid-gap states have also been observed in MI/TBI heterostructure\cite{PhysRevB.87.161108}. Note that such many-body effects in SR mean-field theory cannot be captured in the HF calculations. Thus, the SR results show the robustness of the helical edge modes even when many-body charge fluctuation effects are included.

The interplay of strong correlation-induced magnetism and topology from spin-orbit has thus been actively investigated. The interplay of interaction effect on BTI edge modes has been studied on the Bernevig-Hughes-Zhang (BHZ) model\cite{doi:10.1126/science.1133734}. Within Dynamical Mean Field Theory (DMFT)\cite{capone-0, capone-1} and considering interaction and spin-orbit coupling on the \textit{same region} of the lattice in a strip geometry, the existence of topologically protected metallic edge mode has been established in the paramagnetic regime. The BHZ TI model kept in proximity to a \textit{paramagnetic} Mott insulator (MI), has been shown to support metallic edge models along with interesting edge reconstructions\cite{buried-int}. Finally, paramagnetic DMFT solution on interacting square lattice with spin-orbit interactions sandwiched between two Mott insulators have described the nature of metallic state induced in the edge layers of the Mott insulators\cite{PhysRevB.87.161108}. 

\section{Conclusion}
In previous works on MI/TI interfaces, electron interaction, and spin-orbit effects were simultaneously present in the TI region \cite{capone-0, capone-1} . Further, while inhomogeneous DMFT was employed in earlier work, long-range Coulomb interaction effects were not considered, and only paramagnetic DMFT solutions were evaluated. Thus, buried interfaces of genuine TRS band TI and MI could not be studied as the Coulomb potential is crucial in avoiding electronic phase separation. Even in the cases examined, the paramagnetic calculations\cite{buried-int, PhysRevB.87.161108} have prevented any insight into the magnetic nature of the topological edge mode. Thus, the effect of induced magnetism on TRS-TI buried edges has remained open. 

We have studied the interplay between interaction effects and spin-orbit coupling at the MI-TI interface in Lieb lattice heterostructure geometry. The MI has a ferrimagnetic order, while the TI hosts a bulk topological phase. Upon hybridization, solving the Hartree-Fock problem, and self-consistent evaluation of the Coulomb potential, we find that the metallic edge modes survive despite the edges being subject to finite magnetization. The full heterostructure has a common spin quantization axis, and the induced magnetization in the helical edge modes is small.  Due to this, the topological protection survives. In addition, the Hartree-Fock results show that the induced magnetization at the TI interface unit cells causes a spin imbalance in the edge modes. We demonstrate that this imbalance stabilizes a charge current at the interface that is topologically protected. The current can be controlled by tuning the spin-orbit coupling strength and the interface geometry.
We also support the validity of edge modes from Hartree-Fock by using strong coupling paramagnetic slave-rotor theory that captures charge fluctuation.

The Lieb lattice is easily realized in layered transition metal oxides with perovskites structure \cite{material-1,material-2}. We envision a 3d/5d layered perovskite heterostructure as a natural candidate to realize such topologically protected partially spin-polarized charge currents. The tuning of the parameters can be achieved by choosing differing combinations of the 3d (with strong correlations) and 5d (strong spin-orbit coupling) transition metal atoms.  Such controlled current sources can be used in electronics and spintronic devices to reduce heat generation.


\acknowledgments{We acknowledge the usage of VIRGO and NOETHER computational clusters at NISER. A.M. would also like to acknowledge  SERB-MATRICS grant (Grant No. MTR/2022/000636)  from the Science and Engineering Research Board (SERB) for funding.  K.S. would also like to acknowledge  SERB-MATRICS grant (Grant No. MTR/2023/000743)  from the Science and Engineering Research Board (SERB) for funding. }

\appendix  
\counterwithin*{equation}{section} 
\renewcommand\theequation{\thesection\arabic{equation}} 

\section{Hartree-Fock approach}

\subsection{Mean field in k-space}

This section outlines the HF mean-field formalism introduced in section \ref{sec_2}. The interacting Hubbard Hamiltonian is given by
\begin{equation}
   \label{hubbard1}
\mathcal{H}_{int}=U \sum_{i} \hat{n}_{i \uparrow } \hat{n}_{i \downarrow }.
\end{equation}
We employ the mean-field approximation to simplify the two-body interaction term. This involves decoupling the product of number operators as
\begin{equation}
    \hat{n}_{i \uparrow } \hat{n}_{i \downarrow }\approx\left\langle\hat{n}_{i \uparrow } \right\rangle\hat{n}_{i \downarrow }+\left\langle\hat{n}_{i \downarrow } \right\rangle\hat{n}_{i \uparrow }-\left\langle\hat{n}_{i \uparrow } \right\rangle\left\langle\hat{n}_{i \downarrow } \right\rangle.
\end{equation}
Under this approximation, the interaction Hamiltonian takes the following form:
\begin{equation}
\label{mf1}
\begin{aligned}
&\mathcal{H}_{int}=\sum_i \frac{U}{2}[\left\langle n_i \right\rangle n_i-4\left\langle m_i \right\rangle m_i]-\frac{U}{4}[\left\langle n_i \right\rangle^2-4
\left\langle m_i \right\rangle^2],
\end{aligned}
\end{equation}
where $n_i=(n_{i \uparrow}+n_{i \downarrow})$ is the total on-site occupation and $m_i=\frac{1}{2}(n_{i \uparrow}-n_{i \downarrow})$ is the on-site magnetization. The second term in Eq.~\ref{mf1} introduces a global constant energy shift and can, therefore, be neglected as it does not affect the relative energy spectrum of the system. The resulting interacting Hamiltonian in momentum space is 

\begin{equation}
    \label{mf2}
\mathcal{H}_{int}=\sum_{I_{\alpha},k_y}\frac{U}{2}[\left\langle n^{I_{\alpha}}\right\rangle n^{I_{\alpha}}_{k_y}-4\left\langle m^{I_{\alpha}}\right\rangle m^{I_{\alpha}}_{k_y}],
\end{equation}
where $k_y$ is conjugate momenta along the translation invariant direction $y$ and $\alpha$ denotes sites within the unit cell $I$ as shown in Fig.\ref{fig:1}. $\left\langle n^{I_{\alpha}} \right\rangle=\frac{1}{N}\sum_{k_y} \left\langle n^{I_{\alpha}}_{k_y} \right\rangle$ and $\left\langle m^{I_{\alpha}} \right\rangle=\frac{1}{N}\sum_{k_y} \left\langle m^{I_{\alpha}}_{k_y}\right\rangle$ respectively. Here $\left\langle n^{I_{\alpha}} \right\rangle$ and $\left\langle m^{I_{\alpha}} \right\rangle$ are calculated self consistently.\\

\subsection{Long-range Coulomb interaction}

The average electron density of the MI and TI in the heterostructure is determined by the chemical potential $\mu$. The long-range Coulomb interaction between all charges is crucial in controlling the amount of charge transfer across the interfaces. To account for this effect, we solve the Coulomb potentials $\phi^{I\alpha}$ self-consistently at the mean-field level, following the methodology established in the literature \cite{PhysRevB.77.014441, PhysRevB.88.115136}. The Hamiltonian describes the  long-range Coulomb interaction:

\begin{equation}
   \label{mad2}
    H_{LRC}=\sum_{I_{\alpha},{k_y}} \phi^{I_{\alpha}}n^{I_{\alpha}}_{k_y},
\end{equation}
 where, 
 \begin{equation}
\phi^{I_{\alpha}}=\sum_{\beta\neq \alpha} a_M t \frac{\left\langle\hat{n}_{I_{\beta}}\right\rangle-z_{I_{\beta}}}{\left|\vec{r}_{I_{\alpha}}-\vec{r}_{I_{\beta}}\right|}+\sum_{J \neq I,\beta} a_M t \frac{\left\langle\hat{n}_{J_{\beta}}\right\rangle-z_{J_{\beta}}}{\left|\vec{r}_{I_{\alpha}}-\vec{r}_{J_{\beta}}\right|} \nonumber
 \end{equation}
In this equation, $\left\langle\hat{n}_{J_{\beta}}\right\rangle$ represents the electron charge density at the site ($J_{\beta}$) and $a_M$ is the Madelung constant, as defined in the paper. Throughout the calculations, we assume a uniform background charge $z_{I_{\beta},J_{\beta}}=1$. Solving $\phi^{I_{\alpha}}$ self-consistently allows us to incorporate the long-range Coulomb interaction effect on the charge transfer across the MI-TI interface, ensuring that the electrostatic effects are treated consistently within the mean-field framework.

\subsection{Computational details of self-consistency in Hartree-Fock approach}
We performed self-consistent calculations of the mean-field parameters $\left\langle\hat{n}_{i}\right\rangle$ and $\left\langle\hat{m}_{i}\right\rangle$. We set the convergence tolerance in energy to be  $\sim 10^{-5}$. 
Additionally, we fixed the Madelung constant to $a_M=1$ throughout all calculations to ensure a uniform electron density within the bulk on each side of the heterostructure. We have averaged the Coulomb potential within the $\lambda\neq0$ side, leaving two unit cells from each end of the $\lambda\neq0$ side.

\subsection{Derivation of current}
In section \ref{sec_2} of the main text, we briefly outlined the prescription to calculate the current operators. In this section, we provide derivation of current operators. We start be rewriting the Hamiltonian in a slightly convenient notation. 
This Hamiltonian Eq.  \ref{ham1} can be separated into three components:
\begin{equation}
   \mathcal{H}=\mathcal{H}_t+\mathcal{H}_{SO}+\mathcal{H}_{int},
\label{ham3}
\end{equation}
where $\mathcal{H}_t=-t \sum_{\langle i, j\rangle \in \Lambda} \sum_{\sigma} (d_{i \sigma}^{\dagger} d_{j \sigma} +h.c.)$ is the hopping Hamiltonian, $\mathcal{H}_{SO}=i \lambda \sum_{\langle\langle i, j\rangle\rangle\in \Lambda_{\lambda}} \sum_{\sigma, \sigma'}\nu_{i j} (d_{i \sigma}^{\dagger} \sigma_{\sigma \sigma'}^{z} d_{j \sigma'})$ describes the spin-orbit coupling
and $\mathcal{H}_{int}=U \sum_{i\in \Lambda_{U}} n_{i \uparrow} n_{i \downarrow}$ represents the onsite Hubbard interaction .\\
In the mean-field approximation $[n_I,\mathcal{H}_{int}]=0$, simplifying the analysis. The hopping and spin-orbit coupling terms are rewritten in terms of the unit cell index  $I$. The hopping Hamiltonian becomes
\begin{align}
    \mathcal{H}_t=&-t\sum_{I \sigma} [d^{\dagger}_{ I_a\sigma}d_{ I_c \sigma}+d^{\dagger}_{ I_a\sigma}d_{ I_b \sigma}+d^{\dagger}_{(I+\delta_y)_a \sigma}d_{I_c \sigma}\nonumber\\
    &+d^{\dagger}_{ (I+\delta_x)_a  \sigma}d_{I_b \sigma}]+h.c.
\end{align}
The spin-orbit coupling Hamiltonian is expressed as
\begin{align}
     \mathcal{H}_{SO} = &i\lambda \sum_{I\sigma} s d^{\dagger}_{I_c \sigma} d_{I_b \sigma} - i\lambda \sum_{I\sigma} s d^{\dagger}_{I_b \sigma} d_{I_c \sigma} \nonumber\\
     - &i\lambda \sum_{I\sigma} s d^{\dagger}_{I_c \sigma} d_{(I+\delta_y)_b \sigma} + i\lambda \sum_{I\sigma} s d^{\dagger}_{(I+\delta_y)_b \sigma} d_{I_c \sigma} \nonumber\\
     - &i\lambda \sum_{I\sigma} s d^{\dagger}_{(I+\delta_y-\delta_x)_b \sigma} d_{I_c \sigma} + i\lambda \sum_{I\sigma} s d^{\dagger}_{I_c \sigma} d_{(I+\delta_y-\delta_x)_b \sigma}\nonumber \\
     - &i\lambda \sum_{I\sigma} s d^{\dagger}_{(I+\delta_x)_c \sigma} d_{I_b \sigma} + i\lambda \sum_{I\sigma} s d^{\dagger}_{I _b \sigma} d_{(I+\delta_x)_c \sigma} ,
\end{align}
The local Hubbard interaction term is expressed as:
\begin{equation}
\label{eqn:f}
\mathcal{H}_{int}=U\sum_{I _{\alpha}}  n_{I _{\alpha} \uparrow} n_{I _{\alpha} \downarrow},
\end{equation}
where $U$ is the strength of the on-site Hubbard interaction , $s= 1(-1)$ for $\uparrow(\downarrow)$ . $\delta_x $ and $\delta_y$ are the next-nearest lattice translation vector  relative to the unit cell $I$ in the positive $x$ and $y$ direction respectively. Using these Hamiltonians, the current operators are derived by summing over contributions from all unit cells in a specific direction. The current operator in the $y$-direction is found to be
\begin{align}
    j^y_{\sigma}=&\sum_{I} it (d^{\dagger}_{ (I+\delta_y)_a \sigma}d_{ I _c \sigma}-d^{\dagger}_{ I_c \sigma} d_{ (I+\delta_y)_a \sigma}) \nonumber\\
      + \lambda&\sum_{I} s (d^{\dagger}_{ I_c \sigma}d_{ (I+\delta_y)_b \sigma} + d^{\dagger}_{ (I+\delta_y)_b \sigma} d_{ I_c \sigma})\\
      - \lambda&\sum_{I} s (d^{\dagger}_{ I_c \sigma}d_{ (I+\delta_y-\delta_x)_b \sigma} + d^{\dagger}_{ (I+\delta_y-\delta_x)_b \sigma} d_{ I_c \sigma})\nonumber,
\end{align}
Similarly, the current operator in the $x$-direction is
\begin{align}
j^x_{\sigma}=&\sum_{I} it (d^{\dagger}_{ (I+\delta_x)_a \sigma}d_{ I_b \sigma}-d^{\dagger}_{ I_b \sigma} d_{ (I+\delta_x)_a \sigma})\nonumber\\
- \lambda&\sum_{I} s (d^{\dagger}_{ I_b \sigma}d_{ (I+\delta_x)_c \sigma} + d^{\dagger}_{ (I+\delta_x)_c \sigma} d_{ I_b \sigma})\\
+ \lambda&\sum_{I} s (d^{\dagger}_{ I_b \sigma}d_{ (I+\delta_x-\delta_y)_c \sigma} + d^{\dagger}_{ (I+\delta_x-\delta_y)_c \sigma} d_{ I_b\sigma})\nonumber.
\end{align}

These expressions are derived under the assumption that adjacent layers exist on both sides of the layer that is being considered for the current calculation, i.e., for bulk layers.

\section{Slave-rotor mean field theory}
We study the correlation effects in the heterostructure using the slave-rotor mean-field theory, as outlined in previous works~\cite{PhysRevB.70.035114,PhysRevB.76.195101,PhysRevB.100.045420}. To make this study self-contained, we briefly summarize the method here.
The electronic creation and annihilation operators are decomposed into a direct product of a bosonic rotor degree of freedom and an auxiliary fermion. The rotor accounts for charge occupations, while the spinons preserve the antisymmetric nature of the electronic operators.
To distinguish between the two sides of the heterostructure, we denote the electronic operator on the $U$-side as $c_{I_{\alpha},\sigma}$ and on the $\lambda$-side as $d_{I_{\alpha},\sigma}$. Specifically, the transformations are: 
\begin{align}
    &d_{i,\sigma} \rightarrow c_{I_{\alpha},\sigma} \text{    for  U side } \nonumber,\\
    &d_{i,\sigma} \rightarrow d_{I_{\alpha},\sigma} \text{    for } \lambda\text{  side } .
\end{align}
We then apply the slave-rotor decomposition to the $U$-side as follows:
\begin{equation}
\label{eqn:g}
\begin{array}{l}
c_{I _{\alpha} \sigma}^{\dagger}=f_{I_{\alpha} \sigma}^{\dagger} e^{-i \theta_{I _{\alpha}}}, \\
c_{I _{\alpha} \sigma}=f_{I _{\alpha} \sigma} e^{i \theta_{I_{\alpha}}},
\end{array}
\end{equation}
where $f_{I _{\alpha} \sigma}^{\dagger}$, $f_{I_{\alpha} \sigma}$ are spinon creation and annihilation operators and $e^{\pm i \theta_{I_{\alpha}}}$ represent the rotor creation and annihilation operators.  The rotor operators act on charge states as
\begin{equation}
\label{eqn:h}
e^{\pm i \theta_{I _{\alpha}}}\left|n_{I _{\alpha}}^{\theta}\right\rangle=\left|n_{I _{\alpha}}^{\theta} \pm 1\right\rangle.
\end{equation}
To ensure that the spin and the charge degrees of freedom add up to physical electron occupation in the unit cell, we need to restrict the rotor spectrum. At half-filling, the average occupation of every unit cell is three. To remove the unphysical states from the direct product basis, we impose the following constraint equation
\begin{equation}
\label{eqn:i}
\sum_{\alpha}\left(n_{I _{\alpha}}^{\theta}+n_{I _{\alpha} \uparrow}^{f}+n_{I _{\alpha} \downarrow}^{f}-\mathcal{I}\right)=0,
\end{equation}
where the electron number is equal to the spinon number i.e. $n_{I_{\alpha} \sigma}^{f}=n_{I _{\alpha}\sigma}^{e}$ and $\mathcal{I}$ is the identity operator. 

Then, the Hamiltonian Eq.~\ref{ham1} is reformulated in terms of spinon and rotor operators to derive an exact expression under the slave-rotor decomposition. Subsequently, we introduce a mean-field ansatz to approximate the ground state:
\begin{equation}
\label{eqn:j}
|\Psi\rangle=\left|\Psi^{fd}\right\rangle\left|\Psi^{\theta}\right\rangle,
\end{equation}
where the superscript $d$ denotes the collective index for the electronic operators on the $\lambda$-side, $f$ refers to the spinon operator on the $U$-side, and $\theta$ denotes the rotor operator.  The Hubbard interaction term is confined exclusively to the $ U $-side, resulting in the spinon contribution emerging solely from the $ U $-side. The next step is to compute two decoupled Hamiltonians, $H_{f,d} \equiv\left\langle\Psi^{\theta}|H| \Psi^{\theta}\right\rangle$ and $H_{\theta} \equiv\left\langle\Psi^{fd}|H| \Psi^{fd}\right\rangle$. The expressions are:
\begin{widetext}

\begin{equation}
\label{eqn:k}
\begin{aligned}
H_{f,d}=&-\sum_{I \alpha   \beta \sigma} t_{I_\alpha I_\beta \sigma} \left\langle\Psi^{\theta}\left|e^{-i \theta_{I_\alpha}} e^{i \theta_{I_\beta}}\right| \Psi^{\theta}\right\rangle  f_{I_\alpha \sigma}^{\dagger} f_{I_\beta \sigma}
- \sum_{I\alpha   J\beta \sigma}t_{I_\alpha J_\beta \sigma} d_{I_\alpha \sigma}^{\dagger} d_{J_\beta \sigma}
+\sum_{I\alpha\sigma}\phi^{I_\alpha } \Big( f_{I_\alpha \sigma}^{\dagger} f_{I_\alpha \sigma}+ d_{I_\alpha \sigma}^{\dagger} d_{I_\alpha \sigma}\Big)
\\&-\sum_{I\alpha J\beta \sigma} t_{I_\alpha  J_\beta \sigma}\left\langle\Psi^{\theta}\left|e^{-i \theta_{I_\alpha}} \right| \Psi^{\theta}\right\rangle  f_{I_\alpha \sigma}^{\dagger} d_{J_\beta \sigma}-\sum_{I \alpha \sigma}\big(\lambda_{I_\alpha \sigma}+\mu\big) f_{I_\alpha \sigma}^{\dagger} f_{I_\alpha \sigma},
\end{aligned}
\end{equation}

\begin{equation}
\label{eqn:l}
\begin{aligned}
H_{\theta}=&-\sum_{I \alpha  J \beta \sigma}t_{I_\alpha  J_\beta \sigma}\left\langle\Psi^{fd}\left|f_{I_\alpha \sigma}^{\dagger} f_{J_\beta \sigma}\right| \Psi^{fd}\right\rangle  e^{-i \theta_{I_\alpha}} e^{i \theta_{J_\beta}} 
-\sum_{I\alpha  J\beta \sigma}t_{I_\alpha  J_\beta \sigma}\left\langle\Psi^{fd}\left|f_{I_\alpha \sigma}^{\dagger} d_{J_\beta \sigma}\right| \Psi^{fd}\right\rangle  e^{-i \theta_{I_\alpha}}
+U / 2 \sum_{I \alpha} \Big(n_{I_\alpha}^{\theta}\Big)^2
\\&-\sum_{I\alpha\sigma}\big(\lambda_{I_\alpha \sigma}+\frac{U}{2}\Big) n_{I_\alpha}^{\theta},
\end{aligned}
\end{equation}

\end{widetext}
where $\phi^{I_{\alpha}} =\sum_{J,\beta \neq \alpha} a_M t \frac{\left\langle f_{J _\beta \sigma}^{\dagger} f_{J _\beta \sigma}\right\rangle-z_{J_\beta}}{\left|\vec{r}_{I_\alpha}-\vec{r}_{J_\beta}\right|}+\sum_{J,\beta \neq \alpha} a_M t \frac{\left\langle d_{J_\beta \sigma}^{\dagger} d_{J_\beta \sigma}\right\rangle-z_{J_\beta}}{\left|\vec{r}_{I _\alpha}-\vec{r}_{J_\beta}\right|}$
are onsite Coulomb potentials as discussed in Eq.\eqref{mad1}. The two coupled Hamiltonians are solved self-consistently under the constraint equation. The spinon Hamiltonian refers to Eq.\ref{eqn:k} a one-body physics, while the rotor Hamiltonian in Eq.\ref{eqn:l} describes many-body physics which can be solved using cluster mean-field theory. Since the system is inhomogeneous, we solve using multiple unit cell clusters along the $x$-direction and repeat this in the $y$-direction.

\subsection{Observable in Slave-rotor calculation}
We express the single-particle electron Green's function as a convolution of the spinon and rotor Green's functions\cite{sr-mat}. By taking the imaginary part of this reconstructed Green's function, we compute the spectral function. We define the local (on-site) retarded Matsubara Green's function as

\begin{align}
G_{I _{\alpha} \sigma}\left(i \omega_{n}\right)=&-\int_{0}^{\beta} d \tau e^{i \omega_{n} \tau}\left\langle\Psi\left|c_{I _{\alpha}, \sigma}(\tau) c_{I _{\alpha}, \sigma}^{\dagger}(0)\right| \Psi\right\rangle \nonumber \\
=&-\int_{0}^{\beta} d \tau e^{i \omega_{n} \tau}\left\langle\Psi^{f}\left|f_{I _{\alpha} \sigma}(\tau) f_{I _{\alpha} \sigma}^{\dagger}(0)\right| \Psi^{f}\right\rangle \nonumber\\
& \times\left\langle\Psi^{\theta}\left|e^{-i \theta_{I_{\alpha}}(\tau)} e^{i \theta_{I_{\alpha}}(0)}\right| \Psi^{\theta}\right\rangle.
\label{eqn:n}
\end{align}

The spinon correlation function in Eq. \eqref{eqn:n} is
\begin{equation}
\label{eqn:o}
\begin{array}{l}
\frac{1}{2} \sum_{\sigma}\left\langle f_{I _{\alpha} \sigma}(\tau) f_{I _{\alpha} \sigma}^{\dagger}(0)\right\rangle \\
\quad=\frac{1}{2} \sum_{\alpha \sigma}\left|\left\langle\chi_{\alpha}^{f} \mid I _{\alpha}, \sigma\right\rangle\right|^{2}\left[1-n_{f}\left(\epsilon_{\alpha}^{f}-\mu_{f}\right)\right] e^{-\tau\left(\epsilon_{\alpha}^{f}-\mu_{f}\right)},
\end{array}
\end{equation}
where $\left|\chi_{\alpha}^{f}\right\rangle$ and $\epsilon^f_{\alpha}$ are the spinon eigenvectors and eigenvalues, respectively. The rotor correlation function in Eq. \eqref{eqn:n} is expressed as
\begin{equation}
\label{eqn:p}
\begin{array}{l}
\left\langle e^{-i \theta_{I _\alpha, \sigma}(\tau)} e^{i \theta_{I_\alpha, \sigma}(0)}\right\rangle \\
=\frac{1}{Z_{\theta}} \sum_{m, n} e^{-\beta \epsilon_{m}}\left\langle m\left|e^{-i \theta_{I_\alpha, \sigma}}\right| n\right\rangle\left\langle n\left|e^{i \theta_{I_\alpha, \sigma}}\right| m\right\rangle e^{\tau\left(\epsilon_{m}-\epsilon_{n}\right)},
\end{array}
\end{equation}
where $\epsilon_m$ and $|m\rangle$ are the eigenvalues and corresponding eigenvectors of the rotor Hamiltonian, respectively. Here, $Z_{\theta}$ is the rotor partition function
\begin{equation}
\label{eqn:q}
Z_{\theta}=\sum_{m} e^{-\beta \epsilon_{m}}.
\end{equation}
The integration in Eq. \eqref{eqn:n} performed over imaginary time $\tau$. We then analytically continue back to the real frequency to obtain $G_{I \alpha \sigma}(\omega)$. The projected density of state (PDOS ) is obtained from its imaginary part of the full Green function.

\bibliography{het_bib}
\end{document}